\documentclass[rnote]{aa}

\bibpunct{(}{)}{;}{a}{}{,}

\usepackage{graphicx}
\usepackage{txfonts}

\begin{document} 

\title{Diffusion-desorption ratio of adsorbed CO and CO$_2$ on water ice.}
\titlerunning{Diffusion-desorption ratio of CO and CO$_2$.}
\author{L.J. Karssemeijer\inst{1}
        \and
        H.M. Cuppen\inst{1}
          }
   \institute{Theoretical Chemistry, Institute for Molecules and Materials, Radboud University Nijmegen,\\ Heyendaalseweg 135, 6525 AJ Nijmegen, The Netherlands.\\
              \email{hcuppen@science.ru.nl}
             }
   \date{Received \ldots; accepted \ldots}
 
\abstract
{
Diffusion of atoms and molecules is a key process for the chemical evolution in star-forming regions of the interstellar medium. Accurate data on the mobility of many important interstellar species is often not available, however, which seriously limits the reliability of models describing the physical and chemical processes in molecular clouds.}
{
Here we aim to provide the astrochemical modeling community with reliable data on the ratio between the energy barriers for diffusion and desorption for adsorbed CO and CO$_2$ on water ices.}
{
To this end, we used a fully atomistic, off-lattice kinetic Monte Carlo technique to generate dynamical trajectories of CO and CO$_2$ molecules on the surface of crystalline ice at temperatures relevant for the interstellar medium.}
{
The diffusion-to-desorption barrier ratios are determined to be 0.31 for CO and 0.39 for CO$_2$. These ratios can be directly used to improve the accuracy of current gas-grain chemical models.}
{
} 
   
   
\maketitle
   
\section{Introduction}
Diffusion on icy dust grains is a fundamental process in the chemical evolution of molecular clouds~\citep{Tielens2010,Vidali2012,Herbst2014}. The main reason is that many of the key molecular species are believed to be formed through the diffusive Langmuir-Hinshelwood mechanism on dust grain mantles. At low temperatures, the surface chemistry is dominated by hydrogenation reactions, while at higher temperature, larger species become mobile and lead to the formation of complex organic molecules. In both domains, diffusion is often the rate-limiting step. Furthermore, diffusive processes determine the structure of the ice and lead to trapping of molecular species. Hence, diffusion strongly influences the conditions under which species are released back into the gas phase as the cloud collapses. 

Despite its critical importance, diffusion is a poorly understood process in the field of astrochemistry. As the amount of chemical complexity in modern gas-grain simulation codes continues to grow, this lack of knowledge is becoming an increasingly serious bottleneck that limits the accuracy of the models.

The most recent astrochemical models are capable of simultaneously modeling the chemistry in the gas-phase, which is relatively well understood, and on the grain-surfaces, which is treated by a (possibly multilayered) stochastic or rate-equation method~\citep{Vasyunin2013,Chang2014,Garrod2011,Taquet2012}. In these models, each species $i$ is assigned a binding energy to treat desorption, $E_{\text{bind},i}$, and an energy barrier for grain-surface diffusion, $E_{\text{diff},i}$. Usually, the desorption energies are relatively well defined (at least for the stable species) from either experiment or theory, but this is not the case for the diffusion energy barriers. Because reliable data are often not available, most models assume the diffusion energy to be a universal, fixed fraction, $f$, of the desorption energy: $E_{\text{diff},i} = f E_{\text{bind},i}$. 

The use of this fraction is a key limitation for the models, first and foremost because there is no fundamental physical argument for such a universal ratio to even exist. Instead, this ratio depends on the diffusing species, the substrate, the surface coverage, and possibly on temperature. For this reason the fraction $f$ is very poorly constrained and values between 0.3 and 0.8 are used by the modeling community~\citep{Hasegawa1992,Ruffle2002,Cuppen2009}. As shown by \cite{Vasyunin2013}, however, the value of $f$  seriously influences the outcome of the models. Secondly, from a microscopic point of view, there are also obvious problems with the concept of using a single diffusion and a single desorption barrier because they both vary strongly from site to site, especially in the amorphous ices in the interstellar medium~\citep{Karssemeijer2014}. Inclusion all these local chemical details is not possible for current models, however, and this might not even be necessary
because they aim to provide a more macroscopic view. For this purpose, a single diffusion barrier per species might well be sufficient and may even be desirable, in view of simplicity. But then, this should be a well-constrained species- and environment-specific value. Especially when considering the efficiency of reactions with an activation barrier, it is crucially important to know whether the diffusion rate is higher or lower than the reaction rate. Finally, from a practical point of view, an accurate value of $f$ is also important because it affects the conditions under which the accretion limit is reached and thereby whether or not modelers should use stochastic models.

The purpose of this research note is to provide the astrochemical modeling community with more accurate data on the energy barriers for surface diffusion and thermal desorption of two of the key astrochemical molecules, CO and CO$_2$, on two forms of crystalline water ice. With these data, the accuracy of the models can be improved by making the ratio $f$ species-specific for at least these two molecules.

\section{Computational methodology}
The diffusion and desorption barriers presented in this note were determined using a computational approach developed by the authors in a recent series of papers. For a detailed description of the methods and force fields, we therefore refer to~\cite{Karssemeijer2012,Karssemeijer2014,Karssemeijer2014a}. Summarizing, the calculations are performed using the adaptive kinetic Monte Carlo (AKMC) technique~\citep{Henkelman2001}, as implemented in the EON software package~\citep{Chill2014}\footnote{\url{http://theory.cm.utexas.edu/eon/}}. This is an off-lattice kinetic Monte Carlo (KMC) technique that combines the atomistic detail from molecular dynamics simulations with the ability of probing long timescales of kinetic Monte Carlo. This makes it specifically suitable for studying the details of diffusive processes under astrochemical conditions. To describe the interactions between the molecules, the same set of force fields is used as in the previous studies. The H$_2$O molecules are described using the TIP4P/2005f~\citep{Gonzalez2011} potential. Interactions between water and CO are described in~\cite{Karssemeijer2014} and interactions with CO$_2$ in~\cite{Karssemeijer2014a}. The desorption barrier for each of the surface-binding sites discovered by the AKMC simulations was calculated with respect to the energy of the structurally relaxed isolated substrate and the CO or CO$_2$ admolecule. Corrections accounting for the quantum mechanical zero point energy contribution to the desorption barriers were applied to all results following the method from Appendix B of~\cite{Karssemeijer2014}, to allow for a direct comparison with experimental values.  

\begin{figure}[h]
\centering
\includegraphics{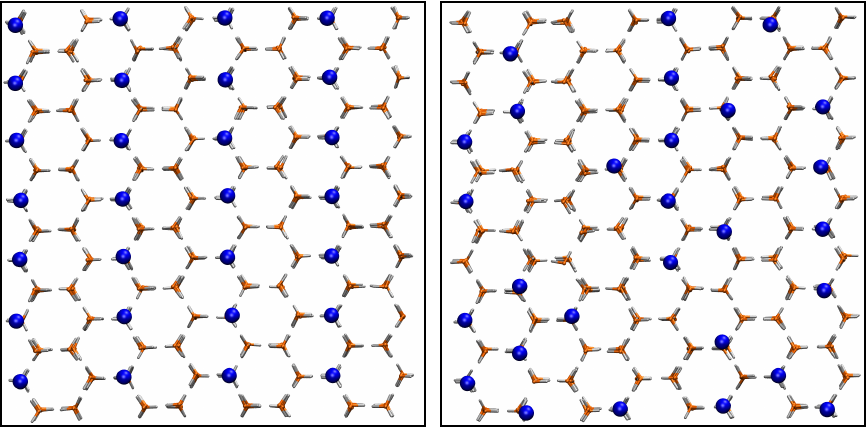}
\caption{Surfaces of the hexagonal ice substrates used in this work. The proton-ordered Fletcher sample is shown on the left, the right panel shows the disordered substrate. Dangling OH bonds are shown in blue.}
\label{fig:samples}
\end{figure}

The surface diffusion of CO and CO$_2$ was studied on the basal plane of two forms of hexagonal ice (ice Ih). The first is the most common form of crystalline ice, which is characterized by a random hydrogen bond network between the oxygen atoms, which sit on tetrahedrally coordinated lattice sites. This partially disordered structure also leads to a random pattern of dangling OH bonds sticking out of the surface, which are known to strongly influence the local binding energy on the surface~\citep{Batista2001,Sun2012}. The second form of hexagonal ice is the so-called Fletcher phase~\citep{Fletcher1992}. This form of ice also has a random hydrogen bond network in the bulk, but has an ordered dangling proton pattern on the surface, with the OH bonds aligned in parallel rows (along the $y$-axis). This ordered structure is believed to minimize the surface energy~\citep{Buch2008,Pan2008} and may therefore be the thermodynamically most stable form of hexagonal ice under typical dense cloud conditions. These two samples are referred to as the `disordered' and the `Fletcher' substrate in the remainder of this paper. The surfaces of the substrates are shown in Fig.~\ref{fig:samples}.

Both samples were created following the procedure given in~\cite{Karssemeijer2014a} and contain 672 H$_2$O molecules with a bulk density of 0.94~g~cm$^{-3}$. They have dimensions of $31\times31$~\AA$^2$, with periodic boundary conditions applied along the $x,y$-plane, parallel to the surface. Because we are only interested in the dynamics of the adsorbed CO or CO$_2$ molecules, the water molecules were constrained from movement, which prevents the morphology of the ice from evolving during the simulation. Because of this constraint, the substrate cannot fully accommodate the admolecules, which typically leads to slightly lower desorption and surface diffusion barriers~\citep{Batista2001,Karssemeijer2014}. The magnitude of this effect was evaluated by also performing simulations on the Fletcher substrate where the topmost 448 H$_2$O molecules were completely free. When the water molecules were allowed to move, we refer to the substrate as `free', when they were constrained from movement, we use the term `frozen'.

\section{Results}

\begin{figure}[h]
\centering
\includegraphics{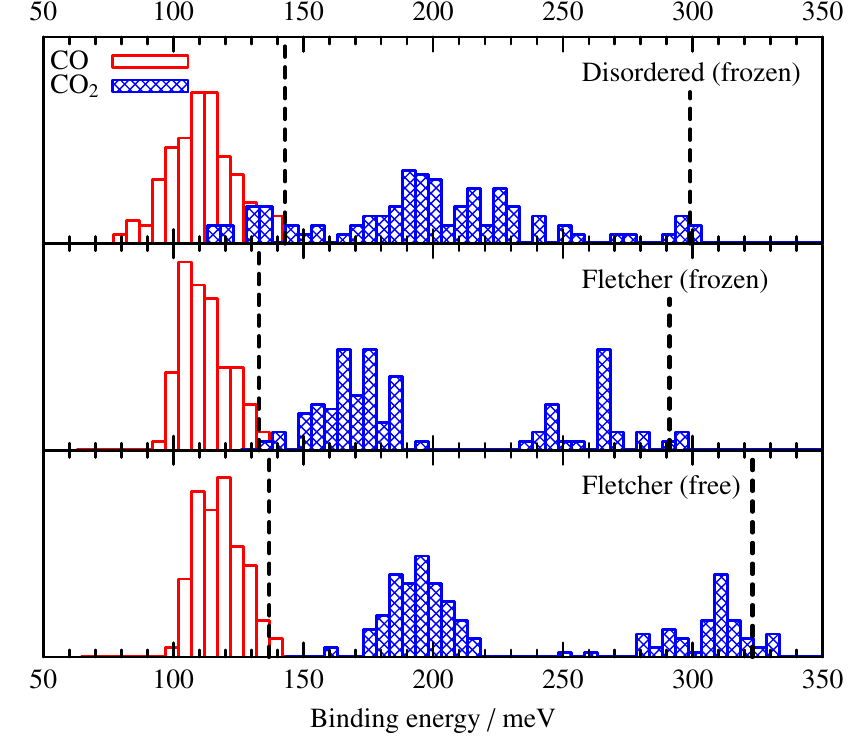}
\caption{Distribution of binding energies of CO and CO$_2$ on the various crystalline ice substrates. The dashed lines indicate the time-averaged binding energies from the kinetic Monte Carlo simulations for CO and CO$_2$ at $T=25$ and $T=70$~K.}
\label{fig:be}
\end{figure}

Using the AKMC scheme, the surface binding sites of CO and CO$_2$ were determined on each of the hexagonal ice substrates. The number of binding sites is listed in Table~\ref{tab:results} and the corresponding distribution of binding energies is shown in Fig.~\ref{fig:be}. The distribution of binding energies is the widest on the proton-disordered substrate for both CO and CO$_2$. This is because the local variation in the arrangement of the dangling OH bonds has a strong influence on the binding energy of small adsorbed molecules. On the Fletcher substrate, the binding sites show more similarity and the distribution of binding energies is sharper. Especially for CO$_2$, two distinct types of binding sites appear on this substrate. These are discussed in detail in \cite{Karssemeijer2014a}. There is still some site-to-site variation, which arises from the disordered hydrogen bond network in the bulk of the ice. As expected, the binding energies on the frozen Fletcher substrate are systematically lower than on the free sample.

\begin{table*}[t]
\caption{Desorption and diffusion energy barriers for CO and CO$_2$ on the proton-disordered and the ordered Fletcher phase of ice Ih.
}
\label{tab:results}
\centering
\begin{tabular}{ll cc c c c c c}
\hline\hline
Substrate & Adsorbate & Binding sites & $E_{\text{bind}}$\tablefootmark{a} &  $E_{\text{diff}}$ & $D_0$ & $E_{\text{diff},x}$ & $E_{\text{diff},y}$ & $f =E_{\text{diff}}/E_{\text{bind}}$ \\ 
 & & & (meV) & (meV) & (cm$^2$ s$^{-1}$) & (meV) & (meV) & \\\hline
Disordered & CO    & 189 & 143  & 49   & 0.051 & 49 & 49 & 0.34\\
(frozen)   & CO$_2$  & 181 & 299  & 127   & 0.018 & 146 & 126 & 0.42\\\hline 
Fletcher   & CO      & 179 & 133  & 39    & 0.025 & 41 & 38 & 0.29\\
(frozen) & CO$_2$  & 169 & 291     & 110    & 0.13 & 150 & 110 & 0.38 \\\hline 
Fletcher & CO      & 182 & 137  & 42  & 0.030 & 42  & 41 & 0.31 \\
(free) & CO$_2$    & 178 & 323  & 122 & 0.23 & 161  & 122 & 0.38\\\hline 
ASW (frozen)\tablefootmark{b} & CO (1+3)\tablefootmark{c} & 96 & 147 & 63 & 0.032 & \ldots & \ldots & 0.42 \\
ASW (frozen)\tablefootmark{b} & CO (1+6)\tablefootmark{c} & 86 & 134 & 48 & 0.009 & \ldots & \ldots & 0.36 \\\hline
\end{tabular}
\tablefoot{The quantities $D_0$ and $E_{\text{diff}}$ are defined in Eq.~\ref{eq:D}\\
\tablefoottext{a}{Binding energies are time-averaged over the kinetic Monte Carlo runs at temperatures of 25 and 70~K for CO and CO$_2$ .}
\tablefoottext{b}{Amorphous solid water (ASW) values are calculated from simulations published in~\cite{Karssemeijer2014} on substrate $S_2^{\text{c}}$. Values are also at $T=25$~K.}
\tablefoottext{c}{There is only one mobile CO molecule on the substrate. The remaining 3 or 6 CO molecules are immobilized in strong binding pore sites on the substrate.}
}
\end{table*}

\begin{figure}[h]
\centering
\includegraphics{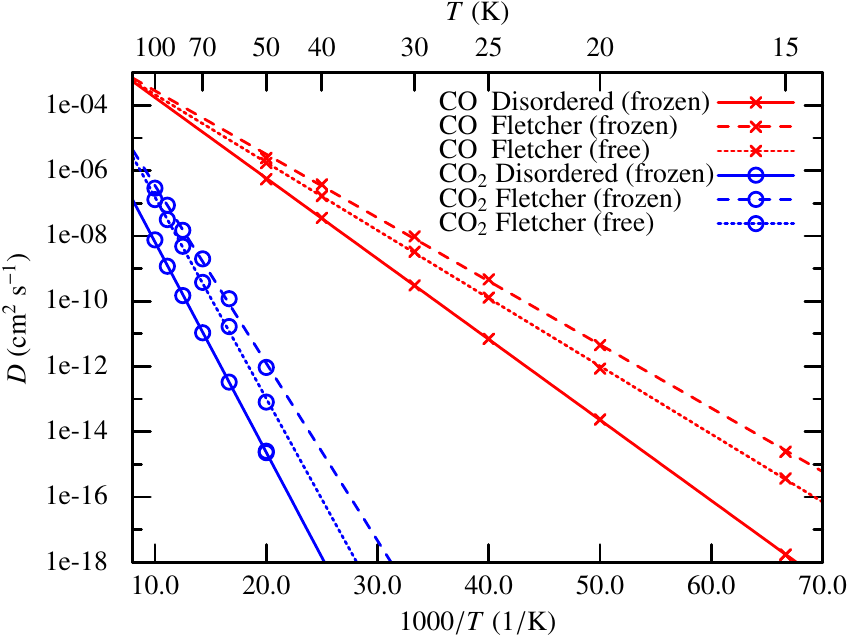}
\caption{Surface diffusion constants for CO and CO$_2$ on the various crystalline ice substrates.}
\label{fig:D_T}
\end{figure}

The mobility of CO and CO$_2$ was studied by generating KMC trajectories at temperatures between 15 and 50~K for CO, and between 50 and 100~K for CO$_2$. From these trajectories, the diffusion constants, $D$, were extracted from the mean squared displacement of the respective admolecule as a function of time~\citep{Frenkel2002}:
\begin{equation}
D =\lim_{t\to\infty} \frac{1}{2dt}  \left< |\vec{r}(t)-\vec{r}(0)|^2\right>.
\end{equation}
Here, $d$ refers to the dimensionality, which is equal to 2 in the case of surface diffusion. As shown in Fig.~\ref{fig:D_T}, the diffusion constants show an Arrhenius behavior as a function of temperature: 
\begin{equation}
D(T) = D_0 \exp{\left(-\frac{E_{\text{diff}}}{k_B T} \right)}.
\label{eq:D}
\end{equation} 
By fitting this expression to the data, the effective activation barrier for diffusion, $E_{\text{diff}}$, and the pre-exponential factor $D_0$ were determined (see Table~\ref{tab:results}).

The KMC trajectories were also used to determine the binding energies. These were time-averaged over the whole simulation at temperatures of 25~K for CO and 70~K for CO$_2$. Because most time is spent in the strongest binding sites, the binding energies are at the edges of the distributions in Fig.~\ref{fig:be}. On the frozen substrates, the strongest binding energies are found on the proton-disordered ice for both CO and CO$_2$. Using these binding energies, the ratio $E_{\text{diff}}/E_{\text{bind}}$ was determined for each adsorbate/surface combination. For CO, this ratio is 0.31 on average, whereas for CO$_2$ a value of 0.39 is found.

The simulations themselves provide very well constrained, reproducible energy barriers for diffusion and desorption, which results in only small uncertainties on the ratio $f$ for a given substrate. When comparing the different crystalline substrates,
however, variations in $f$ on the order of 10\% are observed
because of the different dangling-proton patterns. Compared with the values between 0.3 and 0.8 used in chemical models, however, these variations remain small.

The ordered surface of the Fletcher phase leads to a highly anisotropic surface diffusion. CO and CO$_2$ are both found to diffuse more rapidly along  the $y$-direction, parallel to the dangling OH bonds, than along the perpendicular, $x$, direction. To study this effect, separate analyses were made on the one-dimensional diffusion along these two directions by calculating the one-dimensional mean squared displacements and using Eq.~\ref{eq:D}. The diffusion is found to follow an Arrhenius behavior in one dimension as well, and the activation barriers for the $x$- and $y$-directions are also listed in Table~\ref{tab:results}. For both adsorbates, the activation energies are higher along the $x$- than along the $y$-direction. Similar to the effect on the binding energies,
however, the effect is significantly stronger for CO$_2$ than for CO. For completeness, this analysis was also performed on the disordered structure, where the difference between the energy barriers between the two direction is less than half of the difference on the Fletcher-phase surface. The anisotropy on the disordered substrate arises mainly because of the finite size of the sample, which still introduces some inequality between the $x$- and $y$-directions.

\section{Discussion}
Surface diffusion and desorption are complex processes on the microscopic scale that strongly depend on the local molecular environment. To describe them efficiently in grain-surface chemical codes, however, a simplified treatment is needed. In present-day codes, this simplification is achieved through a global ratio $f=E_{\text{diff}}/E_{\text{bind}}$ that is used to define the diffusion energy barrier based on the binding energy of the specific molecular species. From our calculations, we have determined $f$ for CO and CO$_2$ based on the absolute values of  $E_{\text{diff}}$ and $E_{\text{bind}}$ on crystalline water ice. Although using these species-specific values arguably is an a priori improvement over using a single value for all species, we discuss the validity and assumptions of the calculations below.

The first point to address is the crystalline nature of the water ice substrates we used. Because interstellar ices are believed to be mostly amorphous, the question naturally arises what effect this has on the diffusion to desorption energy ratio. From our previous work on CO dynamics on amorphous solid water (ASW) ice~\citep{Karssemeijer2014}, we know that the mobility of CO is strongly related to the presence of strong binding nanopores on the amorphous ice surface. These sites can have binding energies in excess of 200~meV and can effectively immobilize adsorbed CO molecules. However, these pores do not only increase the effective diffusion energy barrier, but also the average binding energy. Thus, the effect on $f$ of the amorphicity of the substrate will be weaker than the effect on the binding and diffusion energies themselves, compared with crystalline ice. Although a direct calculation of $f$ on ASW at $T=25$~K is computationally not feasible, we calculated the ratio to be 0.42 and 0.36 for CO on amorphous ice where the surface pores are partially occupied with either three or six additional CO molecules (see Table~\ref{tab:results}). 

These values for ASW with an increased adsorbate coverage are not only similar to those on crystalline ice, they may also be more relevant for the interstellar medium than values on the bare ASW substrates. The reason is that the ASW nanopores in molecular clouds are likely to be occupied by species like molecular hydrogen, which is far more abundant and diffuses more rapidly than CO or CO$_2$. In addition, there might not be so many nanopores in interstellar ices because they may simply not survive the long timescales because of pore collapse~\citep{Bossa2012} or because of energy release by exothermic reactions in the pore sites. 

The second point is that in most of the calculations we have made use of frozen substrates, where the water ice itself cannot evolve in time. This simplification effectively lowers the energy barriers for both diffusion and desorption. The diffusion-to-desorption ratio, however, remains largely unaffected, as is clear from comparing the results on the Fletcher substrate, which we simulated in both the free and frozen form.  

This discussion shows that chemical models with a relatively fast diffusion ($f$ between 0.3 and 0.5) are most realistic for describing the reaction rates with CO and CO$_2$. Diffusion of these species is underestimated in models using higher ratios (0.5-0.8). This is hardly surprising because these high ratios are based on the experiments on H$_2$ formation through H atom recombination~\citep{Katz1999a,Perets2005}. For these light atoms, the mechanisms underlying diffusion and desorption may be very different from the small, van der Waals-bonded molecules considered here; especially given the discussion on the possibility of H atom tunneling~\citep{Cazaux2004,Hama2013}. Furthermore, the value of $E_{\text{bind}}$ from \cite{Katz1999a} was given as a lower bound, which means that the ratio $f$ based on these experiments (0.77) is actually an upper bound.

As long as no additional data are available, we recommend using the lower ratios presented here for other small, neutral molecules on water-dominated substrates as well. For radical, light atomic, and charged species, however, the diffusion and desorption mechanisms are most likely very different, and the low ratios we found here may not be appropriate at all. The same holds for diffusion on substrates other than water ice. Here, more research is clearly needed.

\section{Conclusions}
We have presented simulation results on the surface diffusion and desorption energy barriers for CO and CO$_2$ on crystalline water ice. The diffusion-to-desorption barrier ratio is one of the crucial ingredients in current astrochemical gas-grain models, and the accuracy of the models can be significantly improved by using species-specific ratios of 0.31 for CO and 0.39 for CO$_2$. 

These new data close but a small gap of the missing information, and it is up to the astrochemical community to constrain these ratios as well as possible for the other key species and substrates.

\begin{acknowledgements}
This work has been funded by the European Research Council (ERC-2010-StG, Grant Agreement no. 259510-KISMOL), and the VIDI research program 700.10.427, financed by The Netherlands Organization for Scientific Research (NWO). We are thankful for the  stimulating discussions on the so-called $X$-factor ($f$) among the participants of the Lorentz workshop on "Grain-Surface Networks and Data for Astrochemistry" (July 2014, Leiden, The Netherlands). These inspired us to write this note.
\end{acknowledgements}

\bibliographystyle{aa} 

\end{document}